\documentstyle[aps,prd,twocolumn,epsf]{revtex}
\tighten

\begin{document}
\firstfigfalse

\newcommand{\incps}[1]{\centerline{\epsfysize=3.0in \epsfbox{#1}}}
\newcommand{\incxps}[1]{\centerline{\epsfxsize=3.0in \epsfbox{#1}}}


\wideabs{
\title{Causal differencing of flux-conservative equations applied to
black hole spacetimes}

\author{Carsten Gundlach and Paul Walker\\
Max-Planck-Institut f\"ur Gravitationsphysik
(Albert-Einstein-Institut) \\ 
Schlaatzweg 1, 14473 Potsdam, Germany}

\date{4 September 1998}
\maketitle


\begin{abstract}

We give a general scheme for finite-differencing partial differential
equations in flux-conservative form to second order, with a stencil
that can be arbitrarily tilted with respect to the numerical grid,
parameterized by a ``tilt'' vector field $\gamma^A$. This can be used
to center the numerical stencil on the physical light cone, by setting
$\gamma^A=\beta^A$, where $\beta^A$ is the usual shift vector in the
3+1 split of spacetime, but other choices of the tilt may also be
useful. We apply this ``causal differencing'' algorithm to the
Bona-Mass\'o equations, a hyperbolic and flux-conservative form of the
Einstein equations, and demonstrate long term stable causally correct
evolutions of single black hole systems in spherical symmetry.

\end{abstract}


} 

\section{Introduction}

The apparent horizon boundary condition (AHBC) appears to be one of
the fundamental techniques required for evolving black hole spacetimes
using numerical techniques. In this paper, we present work on an AHBC
in the context of the recently formulated Bona-Mass\'o (BM)
hyperbolic system for the Einstein Evolution equations, however in
doing so, we present a technique which is generally causally correct
for {\em any} first order flux-conservative set of PDEs.

The idea of the AHBC was credited to Unruh by
Thornburg\cite{Thornburg87}. The fundamental idea is that, rather than
avoiding a singularity by taking slices which delay the infall of
observers inside the horizon (eventually requiring an infinite force),
one could take regular slices everywhere outside the horizon and place
some stable boundary condition inside the apparent horizon,
excising a portion of the numerical grid. Mathematically this is
consistent because the apparent horizon is known to be inside the
event horizon, and the interior of the event horizon is, by
definition, causally disconnected from its exterior.

The AHBC in numerical relativity was first shown to work by Seidel and
Suen in Ref. \cite{Seidel92a}. The AHBC proposal by Seidel and Suen
has two components.  The first is to choose a shift condition which,
after some evolution, {\em locks} the coordinates by freezing the
position of the horizon and keeping radial distances between points
constant. In spherical symmetry, this uniquely determines the shift
everywhere.  Additionally, to handle large shift terms, Seidel and
Suen propose re-writing the finite difference representation of the
ADM equations to obey the causal structure of the spacetime.

This {\em causal differencing} is an important aspect of much AHBC
work to date and is generally credited to Seidel and Suen (``causal
differencing'') or Alcubierre and Schutz (``causal reconnection'')
\cite{Alcubierre94a}. The idea proposed by Seidel and Suen is as
follows. In the presence of a shift, the Einstein equations have
additional terms in the evolution equations (due to the action of
${\cal L}_\beta$ on $\gamma_{ij}$ and $K_{ij}$). However, there is
another coordinate system in which the shift is zero. Finding the
coordinate system, in general, involves integrating a transformation
function in time. The Seidel and Suen prescription is to finite
difference in the transformed zero shift co-ordinates and then
re-transform the {\em finite difference} representation to coordinates
with a shift.

Using these two techniques, Seidel and Suen proceed to demonstrate
that they can accurately evolve Schwarzschild black holes and
Schwarzschild black holes with infalling scalar fields for long
periods of time.

The followup to Seidel and Suen, a paper by Anninos, Daues, Mass\'o,
Seidel and Suen,
\cite{Anninos94e} gave details of the Seidel and Suen causal
differencing scheme and presented several shift choices.  These shifts
allow for coordinate regularity in the entire spacetime and provide
horizon locking. Moreover, several of these shifts are extensible to
full three dimensional cases, most notably the minimal distortion
shift \cite{York79}. Using causal differencing and horizon locking
shift conditions, Anninos et al.  are able to evolve Schwarzschild
black holes in spherical symmetry for 1000M with very small errors in
the measure of the mass of the horizon, compared to 100\% mass errors
present in simulations without an AHBC around $t=100M$.

The first application of an AHBC to a hyperbolic scheme was that of
Scheel et~al. \cite{Scheel97}. This scheme used a hyperbolic
formulation due to York
on a Schwarzschild black hole. The essence of
the causal difference scheme was to decompose the time derivative into
time evolution along the normal direction and spatial transport due to
the shift. That is, they would evolve along the normal direction to
some point no longer on their numerical grid, and then re-construct
their numerical grid by interpolation. The Scheel et al. approach is roughly
equivalent to our advective (non flux-conservative) causal differencer
with interpolation after the step, as discussed below. The results of
Scheel et~al. were initially disappointing, as an instability arose
on a short ($10 - 100M$) time scale. Very recent work \cite{Scheel98a}
has removed this instability by adding constraints to the evolution
equations (in a manner specific to spherical symmetry), and run times
exceeding $10000 M$ have been achieved.

Anninos et al.  extended their methods by importing a one dimensional
shift into a three dimensional code in \cite{Anninos94c}. They found a
stable horizon mass for a moderate time evolution. This work was
extended in the thesis of Greg Daues \cite{Daues96a}, who applied
various shift and excision conditions to three dimensional systems
evolved in the ADM formulation, successfully evolving a Schwarzschild
hole for 100M using live gauge conditions in three dimensions.

Another demonstration that some aspects of an AHBC could work in
three-dimensional numerical relativity was given by
Br\"ugmann\cite{Bruegmann96}.  Br\"ugmann does not use a shift to
freeze the horizon, and only demonstrates his method in geodesically
sliced spacetimes. That is, the horizon keeps swallowing grid points
as the simulation continues.  Nonetheless, he uses an irregular inner
boundary on a (semi-adaptive) Cartesian grid and demonstrates a
stable evolution for several times longer than the $\pi M$ infall time
of the throat in a geodesically sliced spacetime.  The inner boundary
is simply a few zones inside the location of the (moving) surface
$r=2M$, rather than a numerically located horizon.  The inner boundary
is generated by polynomial extrapolation.

Recently, the Binary Black Hole Grand Challenge has presented several
convincing results using an apparent horizon boundary condition with
an ADM system using a causal differencing scheme similar to that of
Scheel et~al. extended to three dimensions. Using boosted Kerr-Schild
slices of Schwarzschild with exact gauge conditions, the Grand
Challenge was able to transport a black hole across a grid
\cite{Cook97a}. The Grand Challenge also reports the ability to hold a
static black hole static for close to $100M$ in Eddington-Finkelstein
slicings in three dimensions.

The work presented below is an attempt to apply similar techniques to
those used by Daues and by the Grand Challenge to the BM evolution
system, while attempting to exploit the first-order, flux-conservative
form of the BM evolution equations.  In the first half of the paper,
we lay a groundwork for future extensions of the BM system to general
three-dimensional BH spacetimes, which are far beyond the scope of
this paper, by discussing and analyzing various fully
three-dimensional techniques for implementing a general AHBC. In the
second half, we apply the general framework to the BM system applied
to spherically symmetric vacuum spacetimes, that is, to the
Schwarzschild black hole. We evolve initial data taken from three
different time-independent slicings of Schwarzschild, all of which are
regular at the horizon. We use a lapse and shift imported from the
exact solution on all of them, and an exact shift together with a live
harmonic slicing condition on one of them. In each of these situations
we test one traditional and four causal finite differencing schemes.


\section{Causal differencing of flux-conservative
  equations}
\label{sec:causal}

In this section, we develop causal differencing for 
an arbitrary system of flux-conservative equations.  The
equation we want to finite difference is
\begin{equation}
\label{fluxcon1}
{\partial u\over \partial t} + {\partial F^A(u)\over \partial x^A} =
S(u),
\end{equation}
where $x^A$ are the spatial coordinates, and $u$ is a vector of unknowns.
A necessary condition for a stable finite differencing scheme is that
all the characteristics of the system lie inside the numerical domain
of dependence (``the stencil''). 
Depending on the formulation of the
Einstein equations, there may be mathematical characteristics other
than, and in fact outside, the physical light cone. In this paper, we
consider only situations where the mathematical characteristics lie
inside the physical light cone.
In Fig.~\ref{fig:stable_lc} we show the
relationship between the numerical and physical light cone, and show
situations in which stable or unstable situations result including a
causal differencing case, where we adjust the numerical stencil to
follow the physical light cone.  The characteristics are not
immediately apparent from the form (\ref{fluxcon1}) of the equation,
but if we are dealing with a hyperbolic formulation of the Einstein
equations, perhaps coupled to matter, we know that the (physical)
characteristics are centered on the center of the light cones of the
spacetime metric
\begin{equation}
\label{metric1}
ds^2 = - \alpha^2\,dt^2 + g_{AB}(dx^A+\beta^Adt)(dx^B+\beta^Bdt).
\end{equation}
We now introduce an auxiliary coordinate system 
$(\tilde t, \tilde x^a)$ that is related to $(t, x^A)$ by
\begin{equation}
\tilde t = t, \qquad \tilde x^a = \tilde x^a(t,x^A),
\end{equation}
where $\tilde x^a(t,x^A)$ obeys the differential equation
\begin{equation}
\label{define_xtilde}
{\partial \tilde x^a\over \partial t} = \gamma^A {\partial \tilde x^a
\over \partial x^A}.
\end{equation}
The vector field $\gamma^A$ distorts the $\tilde x$ coordinate
system relative to the $x$ coordinate system. 
We define the shorthands
\begin{equation}
m^a \equiv {\partial \tilde x^a \over \partial t}, 
\qquad {M^a}_A \equiv {\partial \tilde x^a\over \partial x^A}, 
\qquad {Q^A}_a
\equiv {\partial x^A \over \partial \tilde x^a}.
\end{equation}
Note that because $\tilde t = t$, $Q$ is also the matrix inverse of
$M$.  


\vbox{
\begin{figure}
\incps{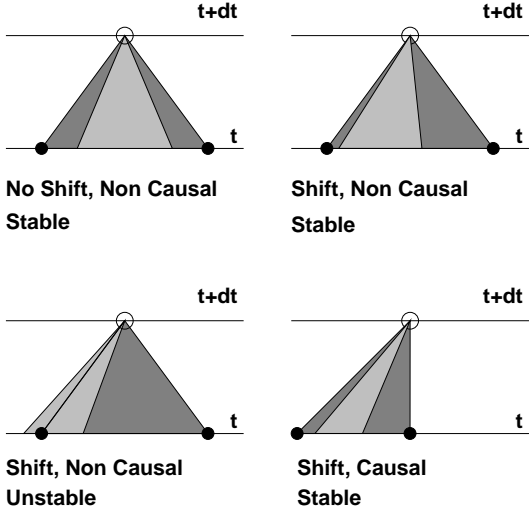}
\caption{We show the stability criterion for light cones and numerical
  stencils. The light gray areas show the physical past light cone of
  a given point, and the dark gray areas show the numerical stencil.
  In the top two figures, where the physical light cone is inside the
  numerical light cone, we will get a stable evolution. In the
  non-causal large shift picture, the evolution will not be stable,
  but tilting the numerical light cone (or causal differencing) will
  result in a stable evolution.}
\label{fig:stable_lc}
\end{figure}
}

Assuming for a moment that we choose
$\gamma^A=\beta^A$, the spacetime metric in the new coordinate system
is 
\begin{equation}
\label{metric2}
ds^2 = -\tilde \alpha^2 d\tilde t^2 + \tilde g_{ab} d\tilde x^a
d\tilde x^b.
\end{equation}
This has no shift, and therefore the light cone is symmetric around
$\partial/\partial\tilde t$.  The 3-metric, lapse and shift
in the two coordinate systems are related by
\begin{equation}
g_{AB} = {M^a}_A {M^b}_B \, \tilde g_{ab}, \qquad 
\alpha = \tilde \alpha, \qquad 
\beta^A = {Q^A}_a m^a.
\end{equation}
Our motivation for introducing the auxiliary $\tilde x$ coordinates
is causal differencing.  Here we take causal differencing to mean
using a stencil that is symmetrically centered on the light cone. This
is equivalent to its being centered on the vector field
$\partial/\partial\tilde t$ in the new coordinate system. We obtain a
``tilted'' stencil in the $x$ coordinates by first transforming the
differential equation to the $\tilde x$ coordinates, using a finite
differencing scheme to advance in $\tilde x$ coordinates, and transforming the
result back to the $x$ coordinates.  Although
$\gamma^A=\beta^A$ is the choice that motivates our scheme, we should
keep in mind the point of view that $\gamma^A$ simply parameterizes
a family of tilted stencils for the differential equation
(\ref{fluxcon1}). No reference needs to be made to either the
spacetime metric or the fact that (\ref{fluxcon1}) is the Einstein
equations. In this respect our scheme differs from the causal
differencing schemes suggested by Seidel and Suen \cite{Seidel92a} 
and Scheel et~al. \cite{Scheel97},
which explicitly use the fact that the Einstein equations simplify in
the $\tilde x$ coordinates. In the following sections we consider
again a generic tilt vector field $\gamma^A$.


\subsection{Transforming the differential equation}

The partial derivatives in the two coordinate systems are related by
\begin{eqnarray}
{\partial \over \partial t} & = & {\partial \over \partial \tilde t} +
m^a {\partial \over \partial \tilde x^a},
\qquad  {\partial \over \partial x^A}  =  {M^a}_A {\partial \over \partial
\tilde x^a}, \\
{\partial \over \partial \tilde t}   & = & {\partial \over \partial
t} - \gamma^A {\partial \over \partial x^A}, \qquad  {\partial \over \partial \tilde x^a }  = {Q^A}_a
{\partial \over \partial x^A}.
\end{eqnarray}
Transforming (\ref{fluxcon1}) to  
$(\tilde t, \tilde x^a)$ we obtain
\begin{equation}
\frac{\partial u}{\partial \tilde t} +
m^a \frac{\partial u}{\partial \tilde x^a} +
{M^a}_A \frac{\partial F^A}{\partial \tilde x^a} =
S(u).
\label{advective}
\end{equation}
We note that this equation is {\em not} in flux-conservative form. We
will refer to (\ref{advective}) as the {\em advective} or {\em
  non flux-conservative} form of our evolution equation. We can put
(\ref{advective}) in flux-conservative form by moving $m^a$ and
${M^a}_A$ into the fluxes, giving additional source terms. Doing this,
we obtain
\begin{equation}
\label{fluxcon2}
{\partial u\over \partial \tilde t} 
+ {\partial \over \partial \tilde x^a}\left(m^a u + {M^a}_A F^A\right) =
S(u) + \lambda \, u + \Lambda_A F^A, 
\end{equation}
where the flux correction coefficients $m^a$ and ${M^a}_A$ have already been
defined, and the source correction terms are
\begin{eqnarray}
\Lambda_A & \equiv & {\partial \over \partial \tilde x^a} {M^a}_A = {\partial \over \partial x^A} \ln \det M, \\
\lambda & \equiv & {\partial \over \partial \tilde x^a} m^a = {\partial \over \partial t} \ln \det M.
\end{eqnarray}
The second equalities have been written down to clarify the function
of the source correction terms, namely to account for a change in the
volume element $\det M$.  It is useful to note the identity
\begin{equation}
\lambda = {\partial
\gamma^A \over \partial x^A} + \gamma^A \Lambda_A.
\end{equation}
We emphasize that when applied to the Einstein Equations, $S(u)$ and
$F^A$ are the BM sources and fluxes {\em including} the shift
terms. Even in the case of $\gamma^A = \beta^A$, we do not cancel
these terms from the sources and fluxes, but allow the cancellation to
take place numerically (both in the advective and flux-conservative
schemes). Below we will explore only the $\gamma^A = \beta^A$ case
numerically, but will find interesting theoretical advantages in the
$\gamma^A = \tau \beta^A$ case (where $\tau$ is some constant).

The advective form of the
equation, (\ref{advective}), has several potential numerical
advantages over the flux-conservative form, (\ref{fluxcon2}). Most
notably, the equation $\partial_t u = 0$ for $u = {\mathrm constant}$
requires cancellations between flux derivatives of the shift and
divergences of the shift in the sources in (\ref{fluxcon2}), while no
such cancellations are needed in (\ref{advective}).

We make the two coordinate systems agree at $t=\tilde
t=0$:
\begin{equation}
\tilde x^a(x^A,0) = {\delta^a}_A x^A.
\end{equation}
In the remainder of this work we also assume that the tilt is
``frozen'' in the $x$ coordinates:
\begin{equation}
{\partial \over \partial t} \gamma^A = 0.
\end{equation}
In practice, the tilt will be constant only throughout one time step,
as described in section~\ref{cd:lapseandshift}.  For convergence tests, we
shall choose it to be constant throughout one time step on the
coarsest grid, and accordingly several time steps on the finer grids.
After each time step we discard the coordinate system $(\tilde t,
\tilde x^a)$ and start again from scratch.  We identify the two
coordinate systems either at the end or at the beginning of the time
step. The implication of this choice on boundary conditions will be
discussed below. We will label the time when the grids coincide as
$t=0$, so time steps will go from $t = -\Delta t$ to $t=0$ or from
$t=0$ to $t=\Delta t$.


\subsection{Calculating the source and flux correction terms}

We need
to know either the $x^A$ traced back to $t=-\Delta t$ along lines of
constant $\tilde x$ or $\tilde x^a$ traced forwards along lines of
constant $x^A$, in order to interpolate the data at the beginning of
the time step onto the $\tilde x$ grid or reconstruct the $x$ grid
after the time step. Rather than by solving the differential equation
(\ref{define_xtilde}) by finite differencing, we do
this by a Taylor series expansion in $t$:
\begin{eqnarray}
\label{Taylor_x}
x^A(\tilde x, t) & = & {\delta^A}_a \tilde x^a - \gamma^A \, t +
{1\over2} \gamma^B
\gamma^A_{,B} \, t^2+ O(t^3), \\
\label{Taylor_tildex}
\tilde x^a(x, t) & = & {\delta^a}_A \left[x^A + \gamma^A \, t +
{1\over2} \gamma^B
\gamma^A_{,B} \, t^2+ O(t^3)\right].
\end{eqnarray}
In order to calculate the corrected fluxes and sources in the
differenced version of (\ref{fluxcon2}), we also need to know $m^a$,
${M^a}_{A}$, $\lambda$ and $\Lambda_A$ on the $\tilde x^a$ grid, at
some of the times $t=\pm\Delta t$, $t=\pm\Delta t/2$ and $t=0$
(exactly which times depend on our numerical integration scheme;
Lax-Wendroff will require the half-times, MacCormack will require the
full times, and so forth). [For (\ref{advective}) we need $m^a$ and
${M^a}_{A}$.] There are different ways of evaluating
these quantities.  We have chosen to expand all auxiliary fields in a
Taylor series around $t=0$ up to $O(t^2)$. As they are only required
at the values $t=\pm\Delta t$ and $t=\pm\Delta t/2$, this should
result in a scheme converging to second order in $\Delta t$. We obtain
\begin{eqnarray}
\label{Taylor_m}
m^a(\tilde x, t) & = & {\delta^a}_A \gamma^A, \\
\label{Taylor_lambda}
\lambda(\tilde x, t) & = & \gamma^A_{,A}, \\
\label{Taylor_M}
{M^a}_A(\tilde x,t) & = & {\delta^a}_A + {\delta^a}_B \gamma^B_{,A} 
t \nonumber \\
&& 
+{1\over2} {\delta^a}_B \left(\gamma^B_{,C}\gamma^C_{,A} +
E^B_{(1)A} \right) 
t^2 + O(t^3), \\
\label{Taylor_Lambda}
\Lambda_A(\tilde x, t) & = & \Lambda_{(1)A} \, t + {1\over2} \Lambda_{(2)A} \, t^2+ O(t^3),
\end{eqnarray}
where we have used the shorthands
\begin{eqnarray}
E^A_{(1)B} & \equiv & - \gamma^C (\gamma^A_{,B})_{,C},\\
\Lambda_{(1)A} & \equiv &   (\gamma^B_{,B})_{,A}, \\
\Lambda_{(2)A} & \equiv & \gamma^B_{,A} \Lambda_{(1)B}- \gamma^B \Lambda_{(1)A,B}.
\end{eqnarray}
Note that all the Taylor coefficients on the right-hand sides are
evaluated at $\tilde t = t = 0$, where the $\tilde x$ and $x$ grids
coincide. As furthermore $\gamma^A$ is independent of time on points
on the $x$ grid, we can think of the right-hand sides as simply
evaluated on the $x$ grid (at whatever time). For the same reason, one
obtains these quantities on half-points of the $\tilde x$ grid (at any
time) simply by averaging the Taylor coefficients between neighboring
points on the $x$ grid: the result is already accurate to quadratic
order.  Note also that $\lambda$ and $m^a$ are exactly constant along
$\tilde x$ lines.


\subsection{Treatment of the lapse and shift}
\label{cd:lapseandshift}

Now we need to address some issues that arise if the system of
flux-conservative equations under consideration are a subset of the
Einstein equations, in our case the Bona-Mass\'o (BM) evolution equations. 

The BM evolution system considers the shift as a given
function of the space and time coordinates. The lapse can be evolved
as a dynamical variable, in which case the entire system is
hyperbolic, or it can also be treated as a given function of the
coordinates, in which case the system is not hyperbolic. In the latter
case, one can consider the lapse as a given function of the spatial
coordinates that is constant for one time step, then changes
discontinuously to be a new constant function for the next time
step. This constant function can then be a functional of the Cauchy
data at the beginning of the time step. One can, for example, obtain
the lapse and shift by solving elliptic equations for maximal slicing
and minimal distortion gauge.  Such an algorithm has no natural
continuum limit (in time). If one wants to verify convergence of the
numerical algorithm, one must define an artificial continuum limit in
which the lapse and shift change at intervals $\Delta t$ which are
multiples of the time step.  One solves the lapse and shift elliptic
equations every time step on the coarse grid, every other time step on
a grid twice as fine, and so on, obtaining a continuum limit in which
the lapse is constant over finite time intervals. In summary, we have
three ways of treating the lapse and two ways of treating the shift.

\begin{enumerate}
\item{Dynamical lapse in the sense of the Bona-Mass\'o formalism.
    There is a hyperbolic evolution equation for the lapse and its
    spatial derivatives. 
        This includes ``1+log'' and harmonic slicing. We shall
    restrict the term ``dynamical lapse'' to this case.}
  
\item{Non-dynamical lapse or shift that nevertheless depends on 
        Cauchy data in the
    manner just described, for example maximal slicing. We shall call
    this a ``live lapse'' or ``live shift.''}
  
\item{A non-dynamical lapse or shift that is really a given function of
    space and time coordinates, obtained from an analytic solution. This includes
    the Kerr-Schild and related slicings of a single black hole. We
    shall call this an ``exact lapse'' or ``exact shift.''}
\end{enumerate}


\subsection{Finite differencing in the $\tilde x$ coordinates}

With the differential equation transformed, we now have to 
finite difference in the $\tilde x$ grid. This will involve creating a
new computational grid by interpolation, and evolving on that grid
rather than the original grid. In this section, we explain the details of
implementing the scheme.

We excise an irregularly shaped region of the spacetime from the
numerical domain by declaring this set of gridpoints to be
``masked''. It is technically easier to allocate memory, and to loop
over grid points, as if these points were still part of the numerical
domain, and then simply to ignore these points. We do this by
overwriting these points with, for example, flat spacetime, and
setting a flag. In one dimension, this operation gives no real savings
in complexity, but in three dimensions, the code is significantly
simplified by this approach.

It is useful to first consider the no shift case, where causal
differencing reduces to ordinary differencing. In order to update
points neighboring the masked region, we have two options. We can
evolve all unmasked points that depend numerically only on unmasked
points, and then recreate the remaining unmasked points by
extrapolation after the time step. Alternatively, we can evolve all
unmasked points after first having created any masked points they
depend on by extrapolation before the time step. The work of Scheel
et~al. \cite{Scheel97} and the Grand Challenge alliance has used extrapolation
after the time step. We have explored both possibilities, as they can
be implemented with the same code.

We use cubic interpolation and extrapolation.

If there is a shift, it will tend to reduce the amount of
extrapolation needed. Ideally it will turn extrapolation into
interpolation (creating a ``boundary without boundary conditions''),
but in the numerical work discussed here, some extrapolation is often
needed. In the presence of a shift (and therefore a tilt in the
stencil) all grid points, not just those at the boundary, need to be
interpolated to new locations. The interpolation and extrapolation are
dealt with in a single algorithm.  To demonstrate the location of the
grids, in Fig.~\ref{fig:causal_lw} we show a one-dimensional
Lax-Wendroff stencil with our causal differencing approach in the
presence of a tilt vector. The top picture indicates grid placements
when we align grids at the beginning of the step, and the bottom
indicates alignment at the end of the step.

There is one technical difference between the two
alternatives. When interpolating/extrapolating after the time step, we
extrapolate only to unmasked points. Because the points are unmasked, we
know the tilt vector at that point, and therefore we know the
$\tilde x$ coordinate values to which we interpolate when re-creating the $x$
grid. If we interpolate/extrapolate before the time step, we must
extrapolate the tilt vector to masked points in order to find the
location of the $\tilde x$ grid in $x$ coordinates inside the stencil
of the boundary point.


\vbox{
\begin{figure}
\incxps{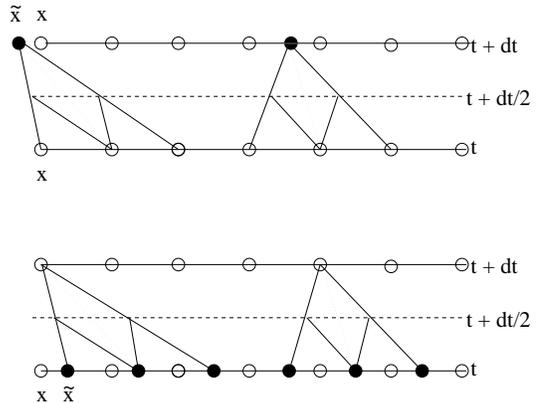}
\caption{
  We show causal Lax-Wendroff stencils, and the
  relationship between the $x$ and $\tilde x$ coordinates for the two
alternatives of interpolation/extrapolation before the time step
(lower diagram) and after the time step (upper diagram). In both
diagrams, the left stencil shows how a ``boundary without boundary
condition'' is achieved if the stencil is tilted enough; The
extrapolation to obtain the leftmost point becomes an
interpolation.}
\label{fig:causal_lw}
\end{figure}
}


Fig.~\ref{fig:cdsetup} illustrates the modular construction of the
causal differencing algorithm. Fig.~\ref{fig:cdsetup} assumes
interpolation before the time step, but an almost identical
prescription exists for interpolation after the time step. At the
beginning of a time step, all fields are known at point A and all
other points on the $x$ grid. We interpolate all dynamical variables
to point C. This is the key step of causal differencing, namely to
interpolate in order to mimic the effect of a transport term in the
evolution equations.
Dynamical variables are $g_{ij}$ and $K_{ij}$, plus in the BM
formalism the $D^i_{jk}$ and $V_i$.  Non-dynamical variables are
$\beta^i$, and in the BM formalism $B^i_j$. $\alpha$ is non-dynamical
in the ADM formalism, while in the BM formalism $\alpha$ and $A_i$ can
be treated either as dynamical or as non-dynamical.


{
\begin{figure}
\incxps{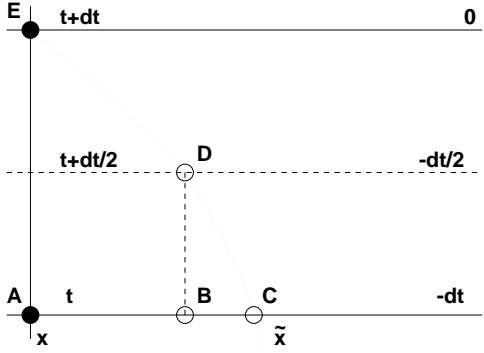}
\caption{Numerical grid showing lines of constant $t$, constant $x^i$, and
  constant $\tilde x^i$. (Recall that $\tilde t = t$.)  Points A and E
  are on the main ($x$) grid, points B, C and D are not. Surfaces of
  constant $t$ are labelled on the left by the coordinate time, and on
  the right by the internal time variable used for Taylor expansions.
  At the beginning of the time step, all fields are given at point A,
  at the end of the time step all fields have been obtained at point
  E. }
\label{fig:cdsetup}
\end{figure}
}

The non-dynamical variables must be provided at points C and D for the
Lax-Wendroff algorithm, and at points C and E for the MacCormack or
MacCormack-like algorithms. If these variables are ``live'' as
described above, this is done by interpolation. Note that live
non-dynamical variables are independent of $t$ along lines of constant
$x$, so that interpolating these variables to point D is the same as
interpolating them to B, and they are the same at E as at A. If the
non-dynamical variables are ``exact'' (derived from an exact
solution), they are set at C, D and E using the correct value of both
$x$ and $t$.

Finally, we need to provide $M^a_A$ and $m^a$ for all causal schemes,
and for the flux-conservative schemes also $\Lambda_A$ and $\lambda$,
at either C and D (for Lax-Wendroff) or C and E (for
MacCormack). These are obtained from the tilt $\gamma^A$ at point E by
a Taylor expansion, as described above. Recall that by assumption
$\gamma^A$ is independent of $t$ along lines of constant $x$,
so that $\gamma^A$ at point E is the same as at point A. 
In
practice $\gamma^i$ is a multiple of the shift $\beta^i$.

If we solve the equations in the $\tilde x$ coordinates in the
flux-conservative form (\ref{fluxcon2}), we can use a standard
evolution algorithm, taking into account only the flux and source
correction terms. Here we have tested a Lax-Wendroff and a MacCormack
algorithm, both incorporating the sources and dealing with all
spatial directions at once. Note that the presence of flux terms in
the source correction terms would make Strang splitting these
equations quite inefficient, and therefore we do not use a Strang
split. 

Since we will investigate both (\ref{fluxcon2}) and
(\ref{advective}) below, it is important to explicitly describe our
numerical method for the advective form, (\ref{advective}).
Differencing the advective form with a MacCormack-like method amounts
to multiplying the finite differences of the fluxes by the appropriate
pre-factor. That is, terms like $\frac{\Delta t}{\Delta x}(F(u_{i+1})
- F(u_i))$ become $\frac{\Delta t}{\Delta x} M (F(u_{i+1}) - F(u_i))$,
where $M$ is the appropriate $m^a$ or ${M^a}_A$ factor. 


\subsection{Testbeds}

We shall test our algorithms on the Schwarzschild spacetime, in a
coordinate system $(t,r,\theta,\phi)$ adapted to spherical
symmetry. As the spacetime is static, there are many coordinate
systems in which all fields are independent of the time
coordinate. Without loss of generality, we use a radial coordinate $r$
defined so that $4\pi r^2$ is the area of any surface $t={\rm const}$,
$r={\rm const}$. In other words, we set $g_{\theta\theta}=r^2$ by
definition. The remaining coordinate freedom is the freedom to slice
the spacetime into surfaces $t={\rm const}$. One slicing in which all
metric coefficients are independent of $t$ is of course $t=t_{\rm
Schw}$, where $t_{\rm Schw}$ is the usual Schwarzschild time
coordinate. This slicing is singular at the event/apparent
horizon. All other slicings which leave the metric coefficients
$t$-independent are of the form $t=t_{\rm Schw} + f(r)$. For one
choice of $f(r)$ [which diverges at the horizon $r=2M$ as
$2M\ln(r-2M)$] one obtains the Eddington-Finkelstein, or Kerr-Schild
slicing, which is regular at the horizon:
\begin{eqnarray}
\alpha^{-2} &=&  g_{rr} = 1+{2M\over r}, \nonumber \\
\beta^r &=& {2M\over r+2M}, \qquad
K_{\theta\theta}=2M\left(1+{2M\over r}\right)^{-1/2}, \nonumber \\
K_{rr}&=&-{2M\over r^2}{r+M\over r+2M}\left(1+{2M\over r}\right)^{1/2}. 
\end{eqnarray}
For another choice of $f(r)$, with the same singular part, but a
different regular part, one obtains the Painlev\'e-G\"ullstrand
slicing, in which the 3-metric is flat:
\begin{eqnarray}
\alpha&=& g_{rr}=1 ,\qquad
\beta^r = \sqrt{2M\over r}, \nonumber \\
K_{rr}&=&-\sqrt{M\over 2r^3}, \qquad K_{\theta\theta}=\sqrt{2Mr}. 
\end{eqnarray}
Finally, we shall consider the harmonic time-independent slicing
\begin{eqnarray} 
\alpha^{-2}&=& g_{rr} = \left(1+{2M\over r}\right)
\left(1+{4M^2\over r^2}\right), \nonumber \\
\beta^r &=& {4 \alpha^2 M^2\over r^2}, \quad
K_{\theta\theta} = {4\alpha M^2\over r}, \nonumber \\
K_{rr} &=& - {4\alpha M^2\over r^3} \left(2 +{3M\over r} + {4M^2\over
r^2} + {4M^3\over r^3}\right),
\end{eqnarray}
which has also been used by Scheel et al. \cite{Scheel98a} Note that
$\alpha^{-2}=g_{rr}$ holds for any time-independent slicing of the
Schwarzschild spacetime.


\subsection{Boundary without boundary conditions}

Fig.~\ref{fig:causal_lw} illustrates a situation in which we can
update grid points at the excision boundary without imposing any
boundary condition and without needing to extrapolate (BWBC). If we
align the $x$ and $\tilde x$ grids before the time step, and the tilt
is so large that the entire stencil lies within the unmasked region,
its base points can be obtained by interpolation. In the absence of a
tilt, we would have to extrapolate into the masked region.  A similar
picture applies when we align the grids after the time step, and one
sees from the two diagrams that the condition on the tilt necessary to
obtain a boundary without boundary condition is the same for the two
alternatives.

In a spherically symmetric situation, both the tilt and the shift are
in the radial direction. A simple calculation then shows how big
the tilt must be to obtain a boundary without boundary condition for a
spherically symmetric black hole. Assuming that both vanish at large
radius, it is sufficiently general to consider a tilt that is a
constant multiple of the shift, that is $\gamma^A = \tau
\beta^A$. We call the constant $\tau$ the ``tilt factor''.

We have to take into account two separate conditions: The tilt must be
large enough to obtain a boundary without boundary condition, and the
physical light cone must lie entirely inside the numerical domain of
dependence as a necessary requirement for numerical stability. Let us
denote the ``Courant number'' $\Delta t/\Delta r$ by $C$, and recall
that $\gamma^r = \tau \beta^r$.  Radial null geodesics
obey
\begin{equation}
\label{lightcone}
{dr\over dt} = -\beta^r\pm{\alpha\over \sqrt{g_{rr}}},
\end{equation}
while the numerical light cones have slopes
\begin{equation}
\label{numcone}
{dr\over dt} = -\tau\beta^r\pm C^{-1}.
\end{equation}
Therefore the conditions that the inner and outer edge of the (past)
physical lightcone lie inside the numerical lightcone are
\begin{eqnarray}
C\tau\beta^r-1 &\le&
C\left(\beta^r-{\alpha\over\sqrt{g_{rr}}}\right)-\delta, \\
C\tau\beta^r+1 &\ge&
C\left(\beta^r+{\alpha\over\sqrt{g_{rr}}}\right)+\delta,
\end{eqnarray}
where $\delta$ is a safety margin (measured in units of $\Delta
r$). It is easy to see that these two conditions are equivalent to
\begin{equation}
\label{Courantcond}
|(\tau-1)\beta^r|\le{1-\delta\over C}-{\alpha\over\sqrt{g_{rr}}}.
\end{equation}
Note that this condition has to be obeyed on the entire grid. The
condition that the inner edge of the numerical light cone is tilted
inwards is easily seen to be
\begin{equation}
\label{bwbc}
\beta^r \ge {1+\epsilon\over C\tau},
\end{equation}
where $\epsilon$ is another dimensionless safety margin. This
condition needs to be obeyed only at the excision radius, and only if
we want to avoid extrapolation.

At large radius, the shift vanishes, while the function
$\alpha/\sqrt{g_{rr}}$ determining the width of the light cone is one
for all usual coordinate systems on Schwarzschild. This gives us the
stability condition $C\le 1-\delta$. In flat spacetime this would be
all. In Kerr-Schild coordinates, we see that $\alpha/\sqrt{g_{rr}}$
decreases from one via $0.5$ at $r=2M$ to zero at $r=0$. If we set
$\tau=1$ (which was done implicitly in previous causal differencing
schemes), the global stability condition is simply $C\le
1-\delta$. The shift grows via $0.5$ at $r=2M$ to one at $r=0$. With
$\tau=1$, it never becomes quite large enough to allow a BWBC. But a
larger value of $\tau$ does allow it, as one can easily see by
plotting the four slopes (\ref{lightcone},\ref{numcone}) against $r$
for different values of $C$ and $\tau$. 

Conversely, for a given finite differencing method one can find the
necessary value of $\tau$ and maximum excision radius $r_0$ to achieve
BWBC. To do this, we take $\epsilon$, $\delta$ and $C$ as given,
assume $\tau\ge1$, set $r=r_0$, saturate the two inequalities
(\ref{Courantcond},\ref{bwbc}), and solve for $\tau$ and $r_0$. For
$\epsilon=\delta=0.2$, $C=0.8$ we obtain $r_0=(2/3)M$ and $\tau=2$. 

In the spatially flat coordinate system, we can generally achieve a
BWBC quite easily, even and typically with $\tau=1$. For
$\epsilon=\delta=0.2$, $C=0.8$ we obtain $r_0=(8/9)M$ and
$\tau=1$. For $\delta=0.3$, $C=0.7$ (larger stability margin) and
$\epsilon=0.05$ (in one dimension we really need no safety margin
here) we obtain again $r_0=(8/9)M$ and $\tau=1$, and this is the case
we have tried numerically.

In general, the natural choice for $\delta$ is $\delta=1-C$, which
makes the stability margin the same at the excision radius as at
infinity. In this case we obtain $\tau=2$ in the Kerr-Schild slicing
and $\tau=1$ in the flat slicing of Schwarzschild, independently of
$\epsilon$.

In three space dimensions in Cartesian coordinates, the shift vector
is not typically aligned with a grid axis, and therefore factors of up
to $\sqrt{3}$ arise in various places. It is clear that BWBC is then
more difficult to achieve. Still, it seems possible for the right
choice of slicing, a tilt factor $\tau>1$, and a sufficiently small
excision radius $r_0$. Excision of black holes in three dimensions
will be investigated elsewhere.


\section{The one-dimensional Bona-Mass\'o system}


\subsection{The equations}

We consider the BM system in spherical symmetry. We choose
coordinates $t$ and $r$ and a diagonal 3-metric. This is the same
system as considered in \cite{Alcubierre97a}, with the addition of a
shift and conformal factor.

\newcommand{\thth}{\theta\theta}
\newcommand{\Brr}{{B_r}^r}

We have four gauge fields,
\begin{equation}
\left(\alpha, A_r, \beta^r, {B_r}^r\right),
\end{equation} 
where $A_r\equiv \alpha_{,r}/\alpha$ and ${B_r}^r\equiv
{\beta^r}_{,r}/2$.  As discussed above, the $\alpha$ and $A_r$ can be
dynamical, live or exact, and the $\beta^r$ and ${B_r}^r$ are only
live or exact.

We have 7 dynamical variables,
\begin{equation}
\left(g_{rr}, g_{\thth}, D_{rrr}, D_{r\thth}, K_{rr}, K_{\thth}, V_r\right),
\end{equation}
where $D_{ijk}\equiv g_{jk,i}/2$ and $V_i\equiv{D_{ij}}^j$. 
We note that $g_{\thth} = g_{\phi\phi} \sin^2 \theta$ by spherical
symmetry. This has the effect of making
\begin{equation}
{D_{rj}}^j = \frac{D_{rrr}}{g_{rr}} + 2 \frac{D_{r\thth}}{g_{\thth}}
\end{equation}
and reducing the definition of $V_r$ to
\begin{equation}
V_r = 2\frac{D_{r\thth}}{g_{\thth}}.
\end{equation}
We also note the useful result
\begin{equation}
K = {K_j}^j = \frac{K_{rr}}{g_{rr}} + 2 \frac{K_{\thth}}{g_{\thth}}
\end{equation}

We optionally introduce a conformal rescaling of the metric, $g
\rightarrow \psi^4 g$, and define $\psi_r = \psi_{,r} / \psi$ and
$\psi_{rr} =  \psi_{,rr} / \psi$. The results in exact spacetimes
given below will {\em not} use a conformal rescaling of the metric,
but we give it for completeness here.

With these choices the non-vanishing BM sources for the Ricci ($n=0$) system
\cite{Alcubierre97a} become
\begin{eqnarray}
S\_\alpha &=& -\alpha^2 f K / \psi^4 + \alpha \beta^r A_r,  \\
S\_{g_{rr}} &=& -2 \alpha K_{rr} / \psi^4 + 4 g_{rr} \Brr + \nonumber \\
 &\,& 2 \beta^r
D_{rrr} + 4 \beta^4 \psi_r g_{rr},
\\
S\_{g_{\thth}} &=& -2 \alpha K_{\thth} / \psi^4 + 2 \beta^r D_{r\thth} + 4
\beta^r \psi_r g_{\thth} \\
S\_K_{rr} &=& 2 K_{rr} \Brr + \alpha \frac{K_{rr}}{\psi^4} \left (2 
  \frac{K_{\thth}}{g_{\thth}} - \frac{K_{rr}}{g_{rr}} \right )
\nonumber \\
 &\,&+
  \alpha A_r \left (\frac{D_{rrr}}{g_{rr}} - 
                    2 \frac{D_{r\thth}}{g_{\thth}}  -2 \psi_r \right) +
                  \nonumber \\
  &\,& 2 \alpha \frac{(D_{r\thth}+2 \psi_r g_{\thth})}{g_{\thth}} \left (
      \frac{D_{rrr}}{g_{rr}} - \frac{D_{r\thth}}{g_{\thth}} \right ) +
    \nonumber \\
  &\,&  2 \alpha A_r V_r +
   8 \alpha A_r \psi_r - \nonumber \\
  &\,&  4 \alpha (A_r \psi_r + \psi_{rr} - \psi_r^2)  \\
S\_K_{\thth} &=& - 2 K_{\thth} \Brr +\alpha \left ( 
  \frac{K_{rr}K_{\thth}}{\psi^4 g_{rr}} - \right.\nonumber \\
 &\,& \qquad \left.
     \frac{(D_{rrr} + 2 \psi_r g_{rr}) (D_{r\thth} + 2 \psi_r
       g_{\thth})}{g_{rr}^2} + 1 \right ) - \nonumber  \\
  &\,&   2 \alpha \left(\frac{A_r \psi_r g_{\thth}}{g_{rr}} +
     \frac{(\psi_{rr}-\psi_r^2) g_{\thth}}{g_{rr}} + \nonumber \right. \\
     &\,& \qquad \left. \frac{2 \psi_r D_{r\thth}}{g_{rr}} -
     \frac{2 \psi_r g_{\thth} D_{rrr}}{g_{rr}^2} \right ) \\
S\_V_r &=& -2 \frac{\alpha}{\psi^4 g_{\thth}} \left [ A_r K_{\thth} -
  \right .\nonumber \\
&\,&\left .
           (D_{r\thth} + 2 \psi_r g_{\thth}) \left (\frac{K_{\thth}}{g_{\thth}} -
                             \frac{K_{rr}}{g_{rr}} \right ) \right ],
\end{eqnarray}
and the non-vanishing fluxes become
\begin{eqnarray}
F\_A_r &=& \alpha f K/\psi^4 - \beta^r A_r  \\
F\_D_{rrr} &=&  \alpha K_{rr}/\psi^4 -2 g_{rr} \Brr \nonumber \\
 &\,& - \beta^r D_{rrr}
   - 2 \beta^r \psi_r g_{rr}, \\
F\_D_{r\thth} &=& \alpha K_{\thth} / \psi^4 - \beta^r D_{r\thth} -
     2 \beta^r \psi_r g_{\thth}, \\
F\_K_{rr} &=& -\beta^r K_{rr} + \alpha \left [ 2 V_r + A_r - 2
  \frac{D_{r\thth}}{g_{\thth}} \right ] \\
F\_K_{\thth} &=& -\beta^r K_{\thth} + \alpha \frac{D_{r\thth}}{g_{rr}}
\\
F\_V_r &=& -\beta^r (V_r + 4 \psi_r).
\end{eqnarray}
All other fluxes and sources vanish identically.
We note that the conformal factor is moved from the fluxes of
the $K_{ij}$ into the sources, and is not finite differenced. $\psi$
and its derivatives are given analytically. 

The Ricci scalar of the 3-metric is 
\begin{eqnarray}
\psi^4 R &=& -4 \frac{\partial_r D_{r\thth}}{g_{rr} g_{\thth}} 
     +4 \frac{D_{rrr} D_{r\thth}}{g_{rr}^2 g_{\thth}} 
     +2 \frac{D_{r\thth}^2}{g_{rr}g_{\thth}^2}
     + \frac{2}{g_{\thth}} \nonumber \\
&\,& - 16 \frac{D_{r\thth} \psi_r}{g_{rr} g_{\theta\theta}}
     - 8 \frac{\psi_{rr}}{g_{rr}}
     + 8 \frac{D_{rrr} \psi_r}{g_{rr}^2}
\end{eqnarray}
and the Hamiltonian constraint,
\begin{equation}
{\cal H} = R + 2 \frac{K_{\thth} (2 K_{rr}g_{\thth} + K_{\thth}
  g_{rr})} {\psi^8 g_{rr} g_{\theta\theta}^2}.
\end{equation}
The maximal slicing equation is
\begin{eqnarray}
\alpha_{,rr} &+& \alpha_{,r} \left [ - \frac{D_{rrr}}{g_{rr}} +
2 \frac{D_{r\thth}}{g_{\thth}} + 2 \psi_r \right ] \nonumber \\
 &=&
\frac{\alpha g_{rr}}{\psi^4} \left [
  \left(\frac{K_{rr}}{g_{rr}}\right)^2 + 2 \left
    (\frac{K_{\thth}}{g_{\thth}} \right ) ^2
\right ].
\end{eqnarray}

\subsection{Numerical results}

\subsubsection{Eddington-Finkelstein}

We present the results of applying our causal differencing schemes to
various black hole spacetimes.  As our base test, we will use an $M=1$
black hole on the domain $1 < r < 4$, so the horizon is at $r=2$ and
our buffer zone has width $1$. We use our excision boundary condition
at the inner boundary, and blend against the analytic solution at the
outer boundary.  We will find, in general, that we can keep the
Eddington-Finkelstein metric stable for many hundreds of $M$ using our
schemes. The error in the solution grows linearly in time, but
converges away {\em faster} than second order towards zero.  In other
words, whereas Scheel et~al. with the unmodified hyperbolic system
\cite{Scheel97} can
achieve run times of order $10M$, using the unmodified BM system, we
can achieve run times in the $100$ to $1000M$ range.

We use the advective and fully flux-conservative
causal MacCormack algorithms. When we interpolate before the step,
we use the analytic value of the shift on the inner boundary point (the only
masked point) to re-locate the stencil for the first un-masked
point. When we interpolate after the step this is not necessary.
We then update the first (and only) un-masked point using an
extrapolator, although this is only for a visual effect, and does not
affect the evolution of the system; with the exception of the shift,
fields could take any value at the innermost (masked) point.

In all cases we report error as either $E = |g_{rr} - g_{rr}^{\rm
exact}| + |g_{\thth} - g_{\thth}^{\rm exact}| + |K_{rr} - K_{rr}^{\rm
exact}| + |K_{\thth} - K_{\thth}^{\rm exact}|$ or as a norm over
Hamiltonian constraint violation. Both properties show the same
convergence behavior. The norm $||$ is the sum of absolute values
divided by the number of grid points.

In Fig.~\ref{fig:nfcmac_ham} we see the Hamiltonian constraint for the
Finkelstein black hole evolved with the advective scheme. The error
grows smoothly until it is of order one, and the code
crashes. Fig.~\ref{fig:comp_1d_errs} below shows that the growth of
the error is between linear and quadratic for the first $200M$ (at the
highest resolution). Later, the growth leading to the crash is
approximately exponential. 

\vbox{
\begin{figure}
\incps{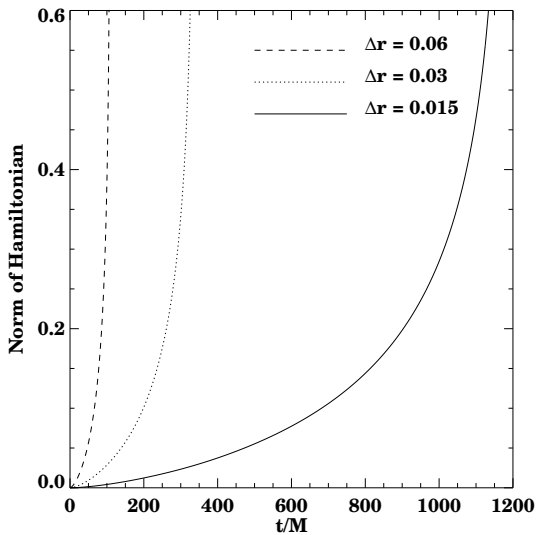}
\caption{We show the evolution of the norm of the Hamiltonian
  constraint for a single Eddington-Finkelstein sliced black hole
  evolved with our advective MacCormack-like scheme. We note that
  generically runtimes are long and errors are small until the code
  crashes. The convergence rate implied by this graph is about 2.7
  until the system crashes. Note that $M=1$ throughout this paper.}
\label{fig:nfcmac_ham}
\end{figure}
}


We repeat this comparison in Fig.~\ref{fig:fullcmac_ham} using the fully
flux-conservative MacCormack scheme, and see the same behavior. That
is, we see that the error converges to zero faster than second order,
and runs are generally cut short by a crash as the error grows
faster-than-linearly towards the end of the run.

\vbox{
\begin{figure}
\incps{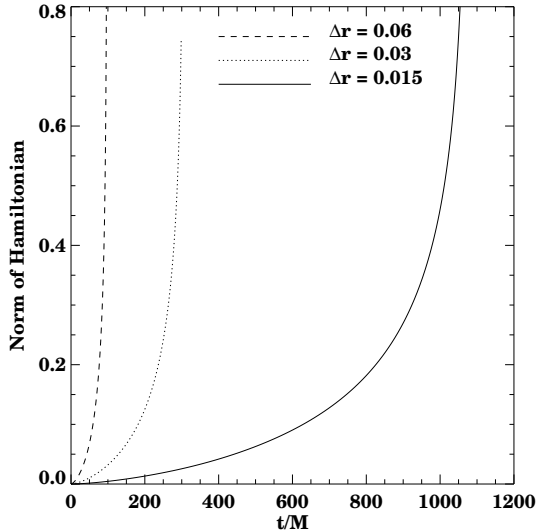}
\caption{We show the Hamiltonian constraint violation at three different
  resolutions for the Eddington-Finkelstein black hole. Compare with
  Fig.~\protect\ref{fig:nfcmac_ham}.}
\label{fig:fullcmac_ham}
\end{figure}
}


Despite this initial transient, the Hamiltonian constraint violation
at late times are essentially the same in both schemes.  In
Fig.~\ref{fig:comp_1d_errs} we compare the error at a fixed resolution
between the two methods, and see exactly this behavior.

\vbox{
\begin{figure}
\incps{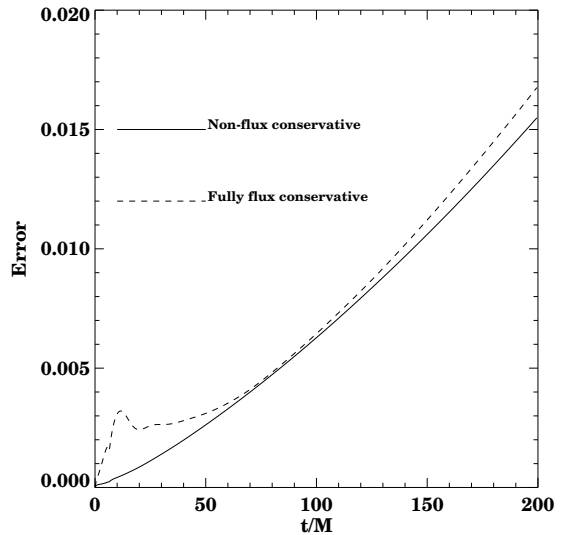}
\caption{We compare the error at $\Delta r = 3/200$ for the fully flux
  conservative and advective methods. We note that, after
  an initial transient error in the fully flux-conservative scheme,
  the errors are essentially the same, and both errors grow in time.}
\label{fig:comp_1d_errs}
\end{figure}
}


The source of the error in our solution is almost entirely dissipation
of the solution at the inner boundary. In Fig.~\ref{fig:grr_1d_plots} we
show plots of $g_{rr}$ on the $\Delta r = 3/200$ grid at times $0M$,
$100M$, and $200M$. We note that the entire error comes from
dissipation at the inner boundary, where the solution slowly ``melts''
away. Our scheme does {\em not} iterate towards a static solution, it
seems, but rather $g_{rr}$ drops linearly (but very slowly) in time,
as shown in the inset.


\vbox{
\begin{figure}
\incps{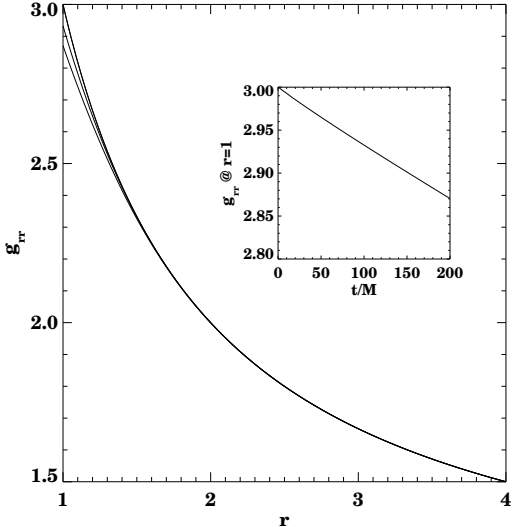}
\caption{We show the evolution of $g_{rr}$ for BM with advective 
  MacCormack using 200 points ($\Delta r = 3/200$). In the large
  figure we show $g_{rr}$ at $0$, $100M$, and $200M$. In the inset we
  show the dissipation of $g_{rr}$ at the inner boundary, which is the
  princiapl source of error.}
\label{fig:grr_1d_plots}
\end{figure}
}


One of the promises of the AHBC is that errors made at the boundary
{\em inside} the horizon won't affect evolution {\em outside} the
horizon. In Fig.~\ref{fig:nfcm_ham} we show this effect. We measure the
Hamiltonian constraint at a given resolution (again, $\Delta r =
3/200$). We observe the constraint converging towards zero at or above
second order, and therefore can be concerned with its
magnitude. Notably, we can see that the violation is several orders of
magnitude larger {\em inside} the horizon. This gives us hope that our
technique is correctly obeying the causal structure of our spacetime;
numerical errors in the exterior are barely affected by (large but
stable) dissipative boundary errors in the interior.


\vbox{
\begin{figure}
\incps{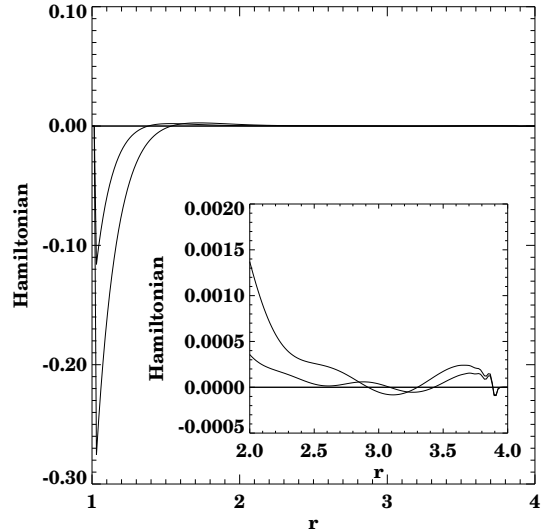}
\caption{We show the Hamiltonian constraint for the BM system with
  the advective causal MacCormack using 200 points
  ($\Delta r = 3/200$) shown at $t=0$, $t=100M$, and $t=200M$. We
  notice that the Hamiltonian constraint violation (which converges to
  zero at second order) is large inside the horizon ($r < 2M$), but in
  the area outside the horizon
  ($r > 2M$) shown in the inset, the violation has not ``escaped.''
  This figure confirms one of the fundamental ideas of the AHBC,
  namely, that an inaccurate but stable scheme applied inside the
  horizon will not affect the system outside the horizon.}
\label{fig:nfcm_ham}
\end{figure}
}


\subsubsection{Spatially flat Schwarzschild, and boundary without a
boundary condition}

We have tested all our numerical schemes on a second slicing of
Schwarzschild, the spatially flat one discussed above. For a direct
comparison with the Eddington-Finkelstein slicing, we have used the
same Courant factor $C=0.5$ and excision radius $r_0=M$. Results are
very similar. Our main observations here are that the finite
differencing schemes again do not tend to a static solution of the
finite difference equations, the error grows approximately linearly in
time, and a doubling of spatial resolution therefore buys a run time
that is four times as long. Comparison with the Eddington-Finkelstein
results confirms the point that our algorithm is not specially made
for a particular solution, works better with higher resolution, and is
therefore generic and robust.  Again, results are much better, at the
same resolution, with causal differencing than without.

In a second series of runs, we have tested a parameter choice, $C=0.7$
and $r_0=0.9M$, that allows us to obtain an excision boundary without
extrapolation. To our knowledge this is the first time that black hole
excision was achieved using causal differencing without extrapolation
at the excision boundary. Nevertheless, the experimental result is
unspectacular: only somewhat larger runtimes are achieved. This
negative result is important as it indicates that extrapolation at the
excision boundary is not the prime cause of error and eventual
instability. Having a no-extrapolation boundary may become more useful
in three space dimensions, where three-dimensional extrapolation on a
jagged ``Lego'' boundary is not as straightforward as in spherical
symmetry.

\subsubsection{Harmonic slicing}

As pointed out by Scheel et al. \cite{Scheel98a}, there is a
time-independent slicing of Schwarzschild that is also harmonic. In
one set of runs, we have used this as our third slicing using an exact
lapse and shift. In a second set of runs, we have evolved the same
initial data with a live lapse, namely the harmonic slicing
condition. With the exact lapse and shift, we find again the same
scaling and run times as for the other two slicings. With the live
harmonic slicing lapse, runtimes at high resolutions are in excess of
$22000M$, much longer than for the exact lapse: here the deviation
from the true solution does not increase monotonously, but turns
around and settles down. As this turnaround occurs at large
deviations, these longer run times are essentially
accidental. Convergence at small errors, however, is still second
order, and the live lapse seems to be as stable as the exact one we
used in the other runs.

In order to obtain a more direct comparison with the results of Scheel
et. al \cite{Scheel98a}, we have made runs with exactly their
resolution, setting the excision radius much closer to the horizon (at
$r=1.8M$, instead of $1.0M$) and/or the outer boundary much further
out (at $r\simeq 120M$ instead of $4.0M$). The combination inner
boundary close to the horizon -- outer boundary close in is
numerically unstable. (We do not know why.) The combination inner
boundary close to the horizon -- outer boundary far out is again
stable. The combination inner boundary at $1.0M$ -- outer boundary far
out works equally well. In summary, in similar circumstances we obtain
similar results as Scheel et al., although the two codes use different
hyperbolic formulations of the Einstein equations.

Our runs are summarized in Table 1.


\section{Conclusions}

Our main result is that while non-causal differencing, with an
extrapolation boundary condition, does not work for black hole
excision, all causal schemes do. The idea of causal differencing is
robust in the sense that all four schemes we have implemented perform
similarly well, and that no modifications of the field equations were
required. In no case does the solution of the finite difference
equations settle down to stable state, so that all runs crash after a
finite time. Nevertheless, the numerical error is as well-behaved as
one can hope for: it is proportional to $h^2$ on the one hand, where
$h$ is the numerical resolution in space and time, and while it is
small, it grows between linearly and quadratically with time on the
other hand. This means, and experience confirms, that with twice the
number of radial grid points one roughly doubles to quadruples the run time
before crashing.

Our results are similar to those of Scheel et
al. \cite{Scheel97,Scheel98a}. In both cases, causal differencing was
applied to a hyperbolic formulation of the Einstein equations. Both
investigations find the same behavior of the numerical error. At the
same resolution, Scheel et al. in their more recent work
\cite{Scheel98a} report run times about a factor of $10$
larger than ours.  One of our four causal differencing schemes
(advective with interpolation at the end) is similar to that of Scheel
et al. The differences are as follows. We have used an exact shift,
and either an exact lapse or the harmonic slicing condition, while
Scheel et al. use the harmonic slicing condition and live minimal
distortion shift. We excise at a fixed coordinate radius, while Scheel
et al. attach their excision radius to the apparent horizon as it
changes through numerical error. In spherical symmetry, these
differences are probably not as important as the finite differencing
scheme itself. 

On physical grounds, no boundary condition is required at the excision
boundary -- it is not a timelike boundary, but a future spacelike
one. (The term ``apparent horizon boundary condition'' is misleading
in this sense, and one better speaks of ``black hole excision''.) We have
used causal differencing to obtain a genuine boundary without boundary
condition, that is, without numerical extrapolation at the excision
boundary. This works well, but does not seem to have a numerical
advantage, at least in spherical symmetry. In other words, the
extrapolation boundary condition does not appear to be the dominant
cause of numerical error in spherical symmetry. (This may well be
different in three space dimensions.)

Our causal differencing methods are immediately applicable to the
Bona-Mass\'o formulation of the Einstein equations in three
dimensions. Ongoing work on black hole excision in three dimensions
will be reported elsewhere.


\acknowledgements

We would like to thank Miguel Alcubierre, Joan Mass\'o, Ed Seidel and
Wai-Mo Suen for interesting and helpful discussions. This work was
supported by the Max-Planck-Institute for Gravitational Physics in
Potsdam.


\bibliographystyle{prsty}
\bibliography{bibtex/references}


\newpage
\widetext
\onecolumn

\begin{table}
\begin{tabular}{cccc|ccc}
\multicolumn{4}{c|}{Method} &
\multicolumn{3}{c}{Runtime in units of $M$} \\
Causal & Form & $\Delta t/\Delta r$ & Interpolate  
& Low & Med & High ($\Delta
r_{\mathrm low} / 4$)\\
\hline
\multicolumn{7}{c}{Eddington-Finkelstein slicing, $\Delta r_{\mathrm
low}=0.06$, $r_0=M$, $r_{\text{max}}=4M$} \\
\hline
 N &FC & 0.5 & --    & 28  & 47 & 95 \\ 
 N &FC & 0.2 & --    &   10  &  16 &  45 \\ 
 N &FC & 0.1 & --    &  7  &  8 & 12 \\ 
 C &Adv& 0.5 & start &  118  &  390 & 1412 \\
 C &Adv& 0.5 & end  & 66 & 170 & 429\\ 
 C &FC & 0.5 & start &  100 &  339 &  1249 \\ 
 C &FC & 0.5 & end  & 58 & 152 & 389 \\ 
\hline
\multicolumn{7}{c}{Spatially flat slicing, $\Delta
r_{\mathrm low}=0.06$, $r_0=M$, $r_{\text{max}}=4M$} \\
\hline
 N &FC & 0.5 & --    & 8.8 & 7.5 & 1.2 \\
 C &Adv& 0.5 & start & 76 & 293 & 1165 \\ 
 C &Adv& 0.5 & end  &  45 & 146 & 451 \\
 C &FC & 0.5 & start & 49 & 293 & 787 \\
 C &FC & 0.5 & end  & 31 & 107 & 339 \\
\hline
\multicolumn{7}{c}{Spatially flat slicing, $\Delta
r_{\mathrm low}=0.06$, $r_0=0.9M$, $r_{\text{max}}=4M$} \\
\hline
 N &FC & 0.7 & --    & 0.6 & 0.3 & 0.2 \\
 C &Adv& 0.7 & start & 127 & 324 & 1411 \\
 C &Adv& 0.7 & end  & 35 & 82 & 503 \\
 C &FC & 0.7 & start & 33 & 26 & 42 \\
 C &FC & 0.7 & end  & 21 & 32 & 33 \\
\hline
\multicolumn{7}{c}{Harmonic slicing, $\Delta
r_{\mathrm low}=0.06$, $r_0=M$, $r_{\text{max}}=4M$} \\
\hline
 N &FC & 0.5 & --    & 16 & 29  & 38 \\
 N &FC & 0.1 & --    & 11 & 13 & 15 \\
 C &Adv& 0.5 & start & 123 & 247 & 863 \\
 C &Adv& 0.5 & end  & 120 & 211 & 688  \\
 C &FC & 0.5 & start & 130 & 251 & 883\\
 C &FC & 0.5 & end  & 137 & 302 & 840 \\
\hline
\multicolumn{7}{c}{Live harmonic lapse, harmonic slicing, $\Delta
r_{\mathrm low}=0.06$, $r_0=M$, $r_{\text{max}}=4M$} \\
\hline
 N &FC & 0.5 & --    & 25 & 51 & 109\\
 N &FC & 0.1 & --    & 15 & 20 & 31  \\
 C &Adv& 0.5 & start & 175 & $>22000 $ & $>22000$  \\
 C &Adv& 0.5 & end  & 121 & $>22000$  & $>22000$ \\
 C &FC & 0.5 & start & 156  & 497 & $>22000$ \\
 C &FC & 0.5 & end  & 116 & 3370 & $>22000$  \\
\hline
\multicolumn{7}{c}{Live harmonic lapse, harmonic slicing, {\bf $\Delta
r_{\mathrm low}=0.125$}, $r_0=1.8M$, $r_{\text{max}}=4M$} \\
\hline
 C &Adv& 0.5 & start & 64 & 68 & 68 \\
 C &Adv& 0.1 & start & 53 & 55 & 55 \\
\hline
\multicolumn{7}{c}{Live harmonic lapse, harmonic slicing, $\Delta
r_{\mathrm low}=0.125$, $r_0=M$, $r_{\text{max}}=120M$} \\
\hline
 C &Adv& 0.5 & start & 16 & 777 & 2103 \\
\hline
\multicolumn{7}{c}{Live harmonic lapse, harmonic slicing, $\Delta
r_{\mathrm low}=0.125$, $r_0=1.8M$, $r_{\text{max}}=120M$} \\
\hline
 C &Adv& 0.5 & start & 68 & 871 & 1389  \\
\end{tabular}
\caption{Summary of black hole evolutions in one dimension. Causal is
either non-causal with extrapolation boundary condition (N) or causal
(C). Form of the equations is either flux-conservative (FC) or
advective (Adv). The integration method is MacCormack or
MacCormack-like (for the advective form). Interpolation is either not
needed (--), or done at the start or the end of the evolution
step. Note that causal advective MacCormack with interpolation after
the time step is essentially the method of Scheel et al. Note that
plain (non-causal) MacCormack, with the same extrapolation boundary
that works for the causal schemes, is unstable in all situations
tried, even for a very small Courant number.}

\label{table:runs}
\end{table}


\end{document}